\documentclass[conference]{IEEEtran}
\usepackage{graphicx} 
\usepackage[hidelinks]{hyperref}
\usepackage{amsmath,graphicx}
\usepackage{amssymb}
\usepackage[space]{cite}

\usepackage{xcolor}
\usepackage{multirow}
\usepackage[acronym,nomain]{glossaries}
\usepackage{subfigure}
\usepackage{booktabs}
\usepackage{color,soul}
\usepackage{threeparttable}
\usepackage[per-mode=symbol]{siunitx}
\usepackage[inkscapelatex=false]{svg}
\usepackage{balance}
\usepackage{array}
\makeglossaries
\newacronym{adc}{ADC}{analog-to-digital converter}
\newacronym{ai}{AI}{Artificial Intelligence}
\newacronym{aoa}{AOA}{angle-of-arrival}
\newacronym{aod}{AOD}{angle-of-departure}
\newacronym{ap}{AP}{access point}
\newacronym{asic}{ASIC}{application specific integrated circuit}
\newacronym{asip}{ASIP}{application specific integrated processor}
\newacronym{awgn}{AWGN}{additive white Gaussian noise}

\newacronym{ce}{CE}{channel estimation}
\newacronym{cw}{CW}{continuous wave}
\newacronym{cdf}{CDF}{cumulative distribution function}
\newacronym{ced}{CED}{cumulative energy density}
\newacronym{clk}{CLK}{clock}
\newacronym{csi}{CSI}{channel state information}
\newacronym{csp}{CSP}{contact service point}
\newacronym{cpu}{CPU}{central processing unit}
\newacronym{crlb}{CRLB}{Cramer-Rao lower bound}
\newacronym{dc}{DC}{direct current}
\newacronym{dlt}{DLT}{Distributed Ledger Technology}
\newacronym{dmimo}{D-MIMO}{distributed massive MIMO}
\newacronym{ram}{RAM}{random-access memory}
\newacronym{ecc}{ECC}{elliptic curve cryptography}
\newacronym{ecdlp}{ECDLP}{elliptic curve discrete logarithm problem}
\newacronym{ecsp}{ECSP}{edge computing service point}
\newacronym{eirp}{EIRP}{equivalent isotropically radiated power}
\newacronym{em}{EM}{electromagnetic}
\newacronym{en}{EN}{energy neutral}
\newacronym{e2e}{E2E}{End-to-End}
\newacronym{pe}{PE}{processing engine}
\newacronym{rtl}{RTL}{register transfer level}
\newacronym{hls}{HLS}{high-level synthesis}
\newacronym{fft}{FFT}{fast fourier transform}
\newacronym{fhss}{FHSS}{frequency hopping spread spectrum}
\newacronym{fl}{FL}{federated learning}
\newacronym{fpga}{FPGA}{Field Programmable Gate Array}
\newacronym{gdpr}{GDPR}{general data protection regulation}
\newacronym{ic}{IC}{integrated circuit}
\newacronym{ioe}{IoE}{Internet of Everything}
\newacronym{iot}{IoT}{Internet of Things}
\newacronym{id}{ID}{information decoding}
\newacronym{kpi}{KPI}{key performance indicator}
\newacronym{lca}{LCA}{life cycle assessment}
\newacronym{ls}{LS}{least squares}
\newacronym{lis}{LIS}{large intelligent surface}
\newacronym{liion}{Li-ion}{lithium-ion}
\newacronym{los}{LoS}{line-of-sight}
\newacronym{nlos}{NLoS}{non-line-of-sight}
\newacronym{lmo}{LMO}{lithium ion manganese oxide}
\newacronym{mf}{MF}{matched filter}
\newacronym{ml}{ML}{machine learning}
\newacronym{mmse}{MMSE}{minimum mean square error}
\newacronym{mmwave}{mmWave}{millimeter wave}
\newacronym{mimo}{MIMO}{multiple-input multiple-output}
\newacronym{miso}{MISO}{multiple-input single-output}
\newacronym{mpc}{MPC}{multipath component}
\newacronym{mrc}{MRC}{maximum ratio combining}
\newacronym{mrt}{MRT}{maximum ratio transmission}
\newacronym{ofdm}{OFDM}{orthogonal frequency-division multiplexing}
\newacronym{ota}{OTA}{over-the-air}
\newacronym{ii}{II}{Initiation interval}
\newacronym{nb}{NB}{narrowband}
\newacronym{nr}{NR}{New Radio}
\newacronym{pa}{PA}{power amplifier}
\newacronym{pdp}{PDP}{power delay profile}
\newacronym{per}{PER}{packet error rate}
\newacronym{pet}{PET}{privacy enhancing technology}
\newacronym{pki}{PKI}{public key infrastucture}
\newacronym{pc}{PC}{pilot count}
\newacronym{pnn}{PNN}{probabilistic neural network}
\newacronym{qrrls}{QR-RLS}{QR decomposition based recursive least squares}
\newacronym{re}{RE}{radio element}
\newacronym{rf}{RF}{radio frequency}
\newacronym{rfid}{RFID}{radio frequency identification}
\newacronym{ris}{RIS}{reflective intelligent surface}
\newacronym{rms}{RMS}{root mean square}
\newacronym{rls}{RLS}{recursive least squares}
\newacronym{rss}{RSS}{received signal strength}
\newacronym{rssi}{RSSI}{received signal strength indicator}
\newacronym{rw}{RW}{RadioWeaves}
\newacronym{sa}{SA}{synchronization anchor}
\newacronym{sdg}{SDG}{Sustainable Development Goal}
\newacronym{sdn}{SDN}{software-defined network}
\newacronym{se}{SE}{spectral efficiency}
\newacronym{sinr}{SINR}{signal-to-interference-plus-noise ratio}
\newacronym{siso}{SISO}{single-input single-output}
\newacronym{slam}{SLAM}{simultaneous localization and mapping}
\newacronym{snr}{SNR}{signal-to-noise ratio}
\newacronym{srls}{SRLS}{standard recursive least squares}
\newacronym{svd}{SVD}{singular value decomposition}
\newacronym{sp}{SP}{service point}
\newacronym{sram}{SRAM}{static random-access memory}
\newacronym{tdd}{TDD}{Time division duplexing}
\newacronym{tdoa}{TDOA}{time-difference-of-arrival}
\newacronym{toa}{TOA}{time-of-arrival}
\newacronym{ue}{UE}{user equipment}
\newacronym{uhf}{UHF}{ultra high frequency}
\newacronym{ula}{ULA}{uniform linear array}
\newacronym{ulp}{ULP}{uplink pilot}
\newacronym{uld}{ULD}{uplink data}
\newacronym{upa}{UPA}{uniform planar array}
\newacronym{ura}{URA}{uniform rectangular array}
\newacronym{uwb}{UWB}{ultrawideband}
\newacronym{wb}{WB}{wideband}
\newacronym{wpt}{WPT}{wireless power transfer}
\newacronym{zf}{ZF}{zero forcing}
\newacronym{Zf}{ZF}{Zero forcing}

\newacronym{lpt}{LPT}{laser power transfer}
\newacronym{rfpt}{RFPT}{radio frequency power transfer}
\newacronym{ipt}{IPT}{inductive power transfer}
\newacronym{cpt}{CPT}{capacitive power transfer}
\newacronym{apt}{APT}{acoustic power transfer}
\newacronym{cfo}{CFO}{carrier frequency offset}
\newacronym{lo}{LO}{local oscillator}
\newacronym{if}{IF}{intermediate frequency}
\newacronym{duc}{DUC}{digital up conversion}
\newacronym{ddc}{DDC}{digital down conversion}
\newacronym{dsp}{DSP}{digital signal processor}
\newacronym{vco}{VCO}{voltage-controlled oscillator}
\newacronym{pll}{PLL}{phase-locked loop}
\newacronym{rt}{RT}{ray-tracing}
\newacronym{lna}{LNA}{low noise amplifier}
\newacronym{dac}{DAC}{digital-to-analog converter}
\newacronym{de}{DE}{drain efficiency}
\newacronym{pae}{PAE}{power-added efficiency}
\newacronym{sar}{SAR}{specific absorption rate}
\newacronym{mppt}{MPPT}{maximum power point tracking}
\newacronym{isi}{ISI}{intersymbol interference}
\newacronym{pps}{PPS}{pulse per second}
\newacronym{ps}{PS}{processing system}
\newacronym{pl}{PL}{programmable logic}
\newacronym{icnirp}{ICNIRP}{International Commission on Non-Ionizing Radiation Protection}
\newacronym{fd}{FD}{Front-haul distance}
\newacronym{wd}{WD}{Wireless distance}
\newacronym{ntp}{NTP}{Network Time Protocol}
\newacronym{ptp}{PTP}{Precision Time Protocol}
\newacronym{wr}{WR}{White Rabbit}
\newacronym{mm2s}{MM2S}{memory mapped to stream}
\newacronym{s2mm}{S2MM}{stream to memory mapped}
\newacronym{ber}{BER}{bit error rate}
\newacronym{mu}{MU}{multiplication unit}
\newacronym{gp}{GP}{general purpose}
\newacronym{hp}{HP}{high performance}
\newacronym{ids}{IDS}{indoor dense space}
\newacronym{dma}{DMA}{direct memory access}
\newacronym{qam}{QAM}{Quadrature amplitude modulation}
\newacronym{qpsk}{QPSK}{Quadrature  phase shift keying}
\newacronym{sd}{SD}{sphere decoding}

\usepackage[printonlyused,withpage]{acronym}

\usepackage{float}
\usepackage{tikz}

\newsavebox{\myimage}

\usepackage{lipsum}  
\newcommand*{\sectionshift}{-0.1cm}

\renewcommand{\vec}[1]{\ensuremath{\boldsymbol{#1}}}
\newcommand{\mrm}[1]{\ensuremath{\mathrm{#1}}}

\IEEEoverridecommandlockouts

\acrodef{ASIC}{application-specific integrated circuit}
\acrodef{ASIP}{application-specific instruction set processor}
\acrodef{BS}{base station}
\acrodef{DMIMO}{distributed massive multiple-input multiple-output}
\acrodef{LIS}{large intelligent surface}
\acrodef{CSI}{channel state information}
\acrodef{ZF}{zero-forcing}
\acrodef{UE}{user equipment}
\acrodef{UL}{uplink}
\acrodef{DL}{downlink}
\acrodef{FFT}{fast fourier transform}
\acrodef{TDD}{time-division duplexing}
\acrodef{FDD}{frequency-division duplexing}
\acrodef{ZF}{zero-forcing}
\acrodef{MMSE}{minimum mean squared error}
\acrodef{MRC}{maximum ratio combining}
\acrodef{BBU}{baseband unit}
\acrodef{SNR}{signal-to-noise ratio}
\acrodef{Tx}{transmit}
\acrodef{Rx}{receive}
\acrodef{AWGN}{additive white Gaussian noise}
\acrodef{MIMO}{multiple-input multiple-output}
\acrodef{OFDM}{orthogonal frequency-division multiplexing}
\acrodef{SIMD}{single instruction, multiple data}
\acrodef{ULP}{\ac{UL} pilot}
\acrodef{ULD}{\ac{UL} data}
\acrodef{CPU}{central processing unit}
\acrodef{LPU}{local processing unit}
\acrodef{LoS}{line of sight}
\acrodef{FPGA}{field programmable gate array}
\acrodef{RF}{Radio Frequency}
\acrodef{GPP}{general purpose processor}

\begin{document}

\title{Spectrum Efficiency and Processing Latency Trade-offs in Panel-Based LIS}

 \author{\IEEEauthorblockN{Lina Tinnerberg, Dumitra Iancu,  Ove Edfors, Liang Liu, and Juan Vidal Alegría  \thanks{This work is funded by the Swedish Foundation for Strategic Research (SSF) project Large Intelligent Surfaces – Architecture and Hardware.}}
 
 \IEEEauthorblockA{Dept. of Electrical and Information Technology, Lund University, Sweden}
 Email: firstname.lastname@eit.lth.se}

	\maketitle

	\begin{abstract}
 The next generation wireless systems will face stringent new requirements, including ultra-low latency, high data rates and enhanced reliability. Large Intelligent Surfaces, is one proposed solution that has the potential to solve these high demands. The real-life deployment of such systems involves different design considerations with non-trivial trade-offs. This paper investigates the trade-off between spectral efficiency and processing latency, considering different antenna distribution schemes and detection algorithms. A latency model for the physical layer processing has been developed, using real FPGA and application-specific instruction processor (ASIP) hardware implementation results. Simulation results using an indoor environment show that distributing antennas throughout the scenario improves overall reliability, while the impact from this on latency is limited both when using zero-forcing (ZF) and Minimum Mean Square Error (MMSE) detection. Changing the detection algorithm to maximum-ratio combining (MRC) from ZF or MMSE, however, reduces the latency significantly, even if a larger number of antennas are needed to achieve a similar spectrum efficiency.

	\end{abstract}

	\glsresetall

	\IEEEpeerreviewmaketitle

	\vspace{\sectionshift}

	\section{Introduction}
	\vspace{\sectionshift}
	\label{sec:intro}

Distributed \ac{MIMO} systems are expected to be a part of the up-and-coming wireless systems that can deliver low-latency, high data-rates and reliable communication. One implementation of such a systems is envisioned through the concept of panel-based active \ac{LIS} with fully digital beamforming,  \cite{8319526}, \cite{9765332}. In this case the LIS consist of a large number of antenna elements with individual coherent transceiver chains which, in this case, are composed of a number of connected panels, each constituting a sub-array. In this configuration, the panels are jointly serving a number of  \acp{UE}. Each panel possesses internal processing resources, as well as inter-panel communication capabilities. 
As the number of antennas increase, so does the amount of data and compute, affecting the total latency of the system and the interconnection bandwidth between antennas and the \ac{CPU}. Decentralized processing approaches available for \ac{LIS} allow overcoming some of the interconnection bottlenecks by distributing the computations throughout \acp{LPU} \cite{9765332}, \cite{8891538}.

Distributed architectures and algorithms for \ac{MIMO} \cite{8849465}, \cite{9414456} and specifically \ac{LIS} \cite{9765332}, \cite{8891538}, have been suggested and analyzed.
In \cite{9765332}, a latency investigation for a panelized LIS system was conducted, with proposed hardware latency models, though without measured latency data from actual hardware. Obtaining real latency numbers from implemented digital signal processing blocks brings valuable insights, especially for understanding practical trade-offs in realistic implementations. These measurements are crucial for defining standards in emerging wireless systems, as they have to meet the performance needs of real-world use cases.

For certain linear detection algorithms in \ac{MIMO} systems, while spectral efficiency improves with the amount of antennas \cite{rusek}, there may still be a large hardware latency overhead, due to the nature and amount of computations that scale with both the number of antennas and the number of users. When comparing linear \ac{MIMO} receivers, their performance heavily depends on the considered scenario. It is known that \ac{ZF}\cite{paulraj} completely eliminates multi-stream interference at the expense of noise enhancement, whereas \ac{MRC} does not handle the interference very well, but works better in low \ac{SNR} scenarios.  \Ac{MMSE} tries to balance between compensating for both noise and interference, but it requires further system assumptions \cite{paulraj}.

This work investigates the trade-offs between performance and latency of the physical layer within a panel-based \ac{LIS} scenario where we scale the \ac{LIS} by adding more panels or further distributing a fixed amount of antennas. We consider the linear processing schemes presented above and we present a possible algorithmic design when distributing the total computing workload onto local panels. Our analysis is based on state-of-the-art implementations of specific hardware for massive \ac{MIMO} linear equalizers. Moreover, we consider the influence of the radio frame structure on the processing latency when the system is serving randomly-placed spatially-multiplexed users.

\begin{figure*}[t]
			\centering 
                \includegraphics[width=1.9\columnwidth]{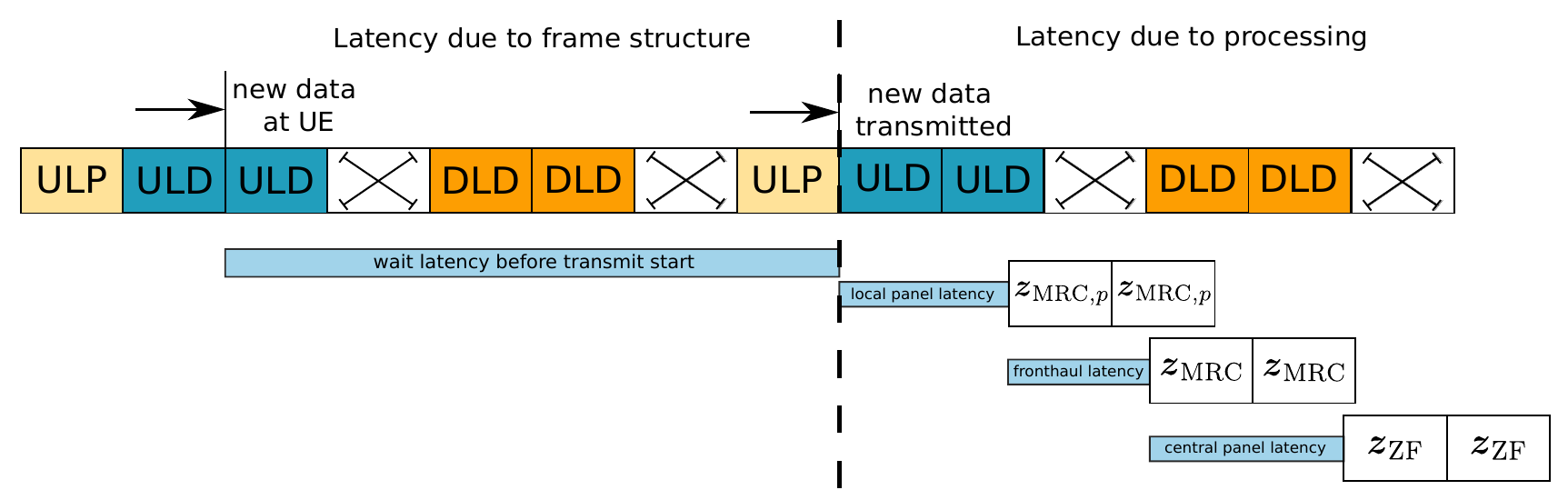}
              \vspace{-0.2cm}
			\caption{A diagram showing the latency of the system and its different sources.}  
            \vspace{-1.5em}
			\label{fig:latency}
		\end{figure*}

\section{System model}

Let us consider an \ac{OFDM}-based system where $K$ single antenna \acp{UE} transmit in the \ac{UL} to a panel-based \ac{LIS}. The panel-based \ac{LIS} consists of $P$ \ac{LIS}-panels, each containing $N$ antennas, so that the whole \ac{LIS} contains a total of $M=NP$ antennas. The vector of symbols received at \ac{LIS}-panel $p\in \{1,\dots,P\}$ within a given subcarrier may be expressed as
\begin{equation}
    \vec{y}_p=\vec{H}_p\vec{s}+\vec{n}_p,
\end{equation}
where $\vec{H}_p$ is the $N\times K$ channel matrix, $\vec{s}$ is the $K\times 1$ vector of symbols transmitted by the \acp{UE} and $\vec{n}_p\sim \mathcal{CN}(0, N_0\mathbf{I}_n)$ corresponds to additive white Gaussian noise. Considering the whole \ac{LIS}, we may express the complete received vector across all P panels as
\begin{equation}
    \vec{y}=\vec{H}\vec{s}+\vec{n},
\end{equation}
where we defined $\vec{y}\triangleq [\vec{y}_1^\mrm{T},\dots,\vec{y}_P^\mrm{T}]^\mrm{T}$,
$\vec{H}\triangleq [\vec{H}_1^\mrm{T},\dots,\vec{H}_P^\mrm{T}]^\mrm{T}$, and $\vec{n}\triangleq [\vec{n}_1^\mrm{T},\dots,\vec{n}_P^\mrm{T}]^\mrm{T}$.

\begin{figure*}[t]
    \centering 
        \includegraphics[width=2\columnwidth]{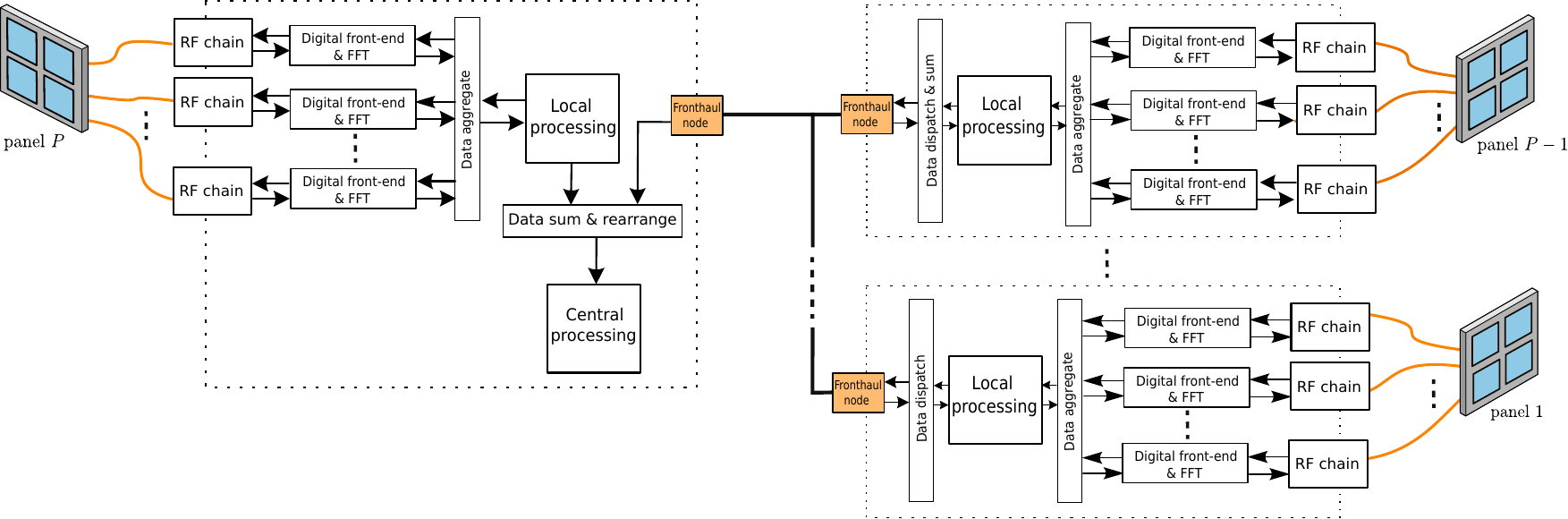}
    \caption{Overview of a general architecture of the processing blocks for panel-based \acp{LIS} connected in a daisy-chain fashion.}  
    \vspace{-1.5em}
    \label{fig:bd}
\end{figure*}

\subsection{Motivation for decentralized processing}
Assuming that all \ac{LIS}-panels are connected to a \ac{CPU}, and that this \ac{CPU} has access to $\vec{y}$, as well as to an estimate of the whole $\vec{H}$,\footnote{In order to have the estimate of $\vec{H}$ at the \ac{CPU} either channel estimation is performed at this \ac{CPU}, or the \ac{LIS}-panels perform channel estimation and forward their estimates of $\vec{H}_p$ $\forall p$ to the \ac{CPU}.} we may coherently equalize the complete received vector by employing common \ac{MIMO} linear processing schemes.\footnote{Linear processing schemes achieve near-optimal performance as the number of base station antennas increase \cite{rusek}.} Three common linear equalizers are \ac{MRC}, \ac{ZF} and \ac{MMSE} \cite{paulraj}, whose equalization matrices are given by
\begin{subequations}\label{eq:lin_proc}
\begin{equation}\label{eq:lin_proc_mrc}
    \vec{W}_\mrm{MRC} = \vec{H}^\mrm{H},
\end{equation}
\begin{equation}\label{eq:lin_proc_zf}
    \vec{W}_\mrm{ZF} = (\vec{H}^\mrm{H}\vec{H})^{-1}\vec{H}^\mrm{H}, \text{ and}
\end{equation}
\begin{equation}\label{eq:lin_proc_mmse}
    \vec{W}_\mrm{MMSE} = (\vec{H}^\mrm{H}\vec{H}+ N_0\vec{I}_K)^{-1}\vec{H}^\mrm{H},
\end{equation}
\end{subequations}
respectively, and the post-processed vector $\vec{z}=\vec{W}\vec{y}$ would be then used to decode user data.

An important limitation of centralized processing is that both the estimate of the whole matrix $\vec{H}$ and the complete received vector $\vec{y}$ have to be known at the \ac{CPU}. This requires sending raw data from the \ac{LIS}-panels to the \ac{CPU}, leading to interconnection bandwidths scaling with number of antennas, $M$, and then performing computations at the \ac{CPU} whose complexity would exhibit the same scaling. A key assumption in systems extending the massive \ac{MIMO} concept, as \ac{LIS}, is to have $M \gg K$ \cite{rusek}, \cite{8319526}. In this case, centralized processing in a distributed system becomes unscalable from the excessive interconnection bandwidth and processing complexity requirements \cite{8849465}. An alternative is to alleviate those requirements by performing part of the processing in parallel at the \ac{LIS}-panels using decentralized processing approaches, as explored next.

We consider decentralized processing approaches where the resulting interconnection bandwidth to a \ac{CPU}, as well as the respective processing complexity, scale with the number of \acp{UE} $K$ instead of with the total number of antennas $M$.\footnote{As argued in \cite{8849465}, the data and resources associated to $K$ simultaneous \acp{UE} inevitably scale with $K$ unless part of the end data is to be ignored.} To this end, each \ac{LIS}-panel $p$ is associated to a \ac{LPU} which can estimate the local channel $\boldsymbol{H}_p$ and perform simple computations based on this estimate. For simplicity we assume that perfect estimates of $\boldsymbol{H}_p$ are obtained, since estimation errors would essentially compromise performance in the same way as for a centralized processing scheme. Moreover, we consider that the \ac{LIS}-panels are sequentially connected via unidirectional links, forming a Daisy chain topology \cite{jesus}. Each link may then be used to share data between panels and  would at most support interconnection bandwidth scaling with $K$. The considered architecture is illustrated in Fig.~\ref{fig:bd}.

\section{Implementation Considerations}

\subsection{Algorithmic design}
If we examine the linear processing schemes given in \eqref{eq:lin_proc}, which assume centralized processing, we may note that the resulting post-processed vectors may be expressed in terms of the local channels by
\begin{subequations}\label{eq:lin_dist}
  \begin{equation}\label{eq:mrc_post}
  \begin{aligned}
\boldsymbol{z}_\mathrm{MRC} &= \sum_{p=1}^{P}\boldsymbol{H}_p^\mathrm{H}\boldsymbol{y}_p
= \sum_{p=1}^{P}\vec{z}_{\mathrm{MRC},p},
\end{aligned}
\end{equation} 
\begin{equation}\label{eq:zf_post}
\begin{aligned}
\boldsymbol{z}_\mathrm{ZF} &=\boldsymbol{G}^{-1}\boldsymbol{z}_\mathrm{MRC}
=\bigg(\sum_{p=1}^{P}\boldsymbol{G}_p\bigg)^{-1}\boldsymbol{z}_\mathrm{MRC},
\end{aligned}
\end{equation}
\begin{equation}\label{eq:mmse_post}
\begin{aligned}
\boldsymbol{z}_\mathrm{MMSE} =\bigg(\sum_{p=1}^{P}\boldsymbol{G}_p+N_0 \boldsymbol{I}_K\bigg)^{-1}\boldsymbol{z}_\mathrm{MRC},
\end{aligned}
\end{equation}
\end{subequations}
where $\vec{z}_{\mathrm{MRC},p}$ is the local MRC-post-processed vector and $\boldsymbol{G}_p=\boldsymbol{H}^\mathrm{H}_p\boldsymbol{H}_p$ is the local Gramian at \ac{LIS} panel $p$.  Hence, we can achieve the same performance as with centralized processing by simply computing at each panel the local MRC-post-processed vector and the local Gramian, and aggregating these throughout the network. Note that $\vec{z}_{\mathrm{MRC},p}$ is a $K\times 1$ vector and $\vec{G}_p$ is an $K\times K$ matrix, so their dimensions scale with the number of \acp{UE}, but not with the total number of antennas.

Considering the Daisy chain interconnection topology, the aggregation of $\vec{z}_{\mathrm{MRC},p}$ and $\boldsymbol{G}_p$ can be done sequentially. Assuming that the \ac{LIS}-panels are numbered according to their position in the Daisy chain, the starting \ac{LIS}-panel $1$ would then share directly  $\vec{z}_{\mathrm{MRC},1}$ and $\boldsymbol{G}_1$ to \ac{LIS}-panel $2$, which would add them with its respective contributions $\vec{z}_{\mathrm{MRC},2}$ and $\boldsymbol{G}_2$ and share the results to the next \ac{LIS}-panel. This would go on sequentially until the final \ac{LIS}-panel $P$ is reached. After adding its contributions, \ac{LIS}-panel $P$ has a complete estimate of both $\vec{z}_\mathrm{MRC}$ and $\vec{G}$, which may be used to perform \ac{ZF} or \ac{MMSE} equalization, i.e., by inverting $\vec{G}$ (or its MMSE regularized version) and multiplying it to $\vec{z}_\mathrm{MRC}$. In this case data shared through each connection would scale quadratically with $K$. On the other hand, if we  restrict processing to \ac{MRC}, there is no need for sharing and aggregating the $\vec{G}_p$ matrices, so the data shared through each connection only scales linearly with $K$. However, we should note that the $\vec{z}_{\mathrm{MRC},p}$ vectors are directly related to transmitted symbols, while the $\vec{G}_p$ matrices are only related to the channel. Thus, for large coherence bandwidths we can reduce the scaling of the data required to share the aggregated $\vec{G}_p$ matrices through each connection. For example, if the coherence bandwidth is large enough so that we can perform channel estimation with a single time-slot, i.e., by having the \acp{UE} transmit their pilots over non-overlapping subcarriers, \ac{MRC}, \ac{ZF}, and \ac{MMSE} would all require interconnection bandwidths scaling linearly with the number of \acp{UE}. In any case, the total number of sequential unidirectional transmissions needed to implement the previous equalizers in a Daisy chain decentralized architecture becomes $P-1$.

\subsection{Reference processing architecture} 

We consider the hardware architecture in Fig.~\ref{fig:bd} for $P$ panels, each having $N$ antennas. 
Each antenna is connected to a processing chain within the \acp{LPU}. The chain begins with the  \ac{RF} stage, which amplifies, filters the analog signal and converts it to the digital domain. It then moves to the digital front-end, where the signal is downsampled, filtered, and transformed via a \ac{FFT} to obtain the frequency-domain signal. The processed signals from all antennas within one panel are aggregated into a matrix form in order to obtain the local MRC-post-processed vector and the local Gramian, given in \eqref{eq:lin_dist}. The data is then prepared to be transferred between panels through the fronthaul infrastructure. For panels $p>1$, the local contributions are summed  with the incoming data from the preceding panels in the Daisy chain. The final panel in the chain acts both as a \ac{LPU} for its antennas, and as a \ac{CPU} for the fully aggregated data. For \ac{MRC}, the equalization is done after the aggregation, and for \ac{ZF} and \ac{MMSE}, matrix inversion of the Gramian (or the \ac{SNR} regularized version for \ac{MMSE}) and subsequent multiplication with the MRC vector is performed in the \ac{CPU}.

The local and central \ac{MIMO} processing can be performed with different types of hardware architectures, ranging from \acp{ASIC} to more generalized architectures such as a \acp{GPP}. It is important to note that performance metrics can vary greatly depending on the chosen architecture. For this paper, we consider one of the processing units implemented in \cite{mj}: a vector core, part of an \ac{ASIP}. It can be leveraged for both the local and central \ac{MIMO} processing, and it is specifically tailored for \ac{MIMO} equalization operations. An \ac{ASIP} is higher performing than a standard CPU, but more flexible than an ASIC, and it can be programmed to execute the processing for different amount of \acp{UE} and antennas\cite{mj}.

\begin{table}[h]
\caption{Latency measured for the \ac{ASIP}\cite{mj} considered with 16 antennas per panel and 16 or 128 \acp{UE}.}

\resizebox{1\columnwidth}{!}{%
\begin{tabular}{@{}lllc@{}}
\toprule
\textbf{Operation} & \multicolumn{2}{l}{\textbf{Latency (clock cycles)}} & \textbf{Complexity}\\ 
       & K=16                     & K=128  & \textbf{order} \\ 
\midrule
$\vec{H} = \vec{y}\vec{S}^{-1}_\text{Pilot}$  &  83         &          531         & $\mathcal{O}(KN)$          \\
$\vec{G} = \vec{H}^\mathrm{H} \vec{H} $        & 946 & 58k & $\mathcal{O}(K^2N)$\\
$\vec{G}^{-1}$                & 1817 & 890k  & $\mathcal{O}(K^3)$\\
$\vec{z}_\mrm{MRC} = \vec{H}^\mathrm{H} \vec{y} $      & 81   & 460                          & $\mathcal{O}(KN)$\\
$\vec{z}_\mrm{ZF} = \vec{G}^{-1} \vec{z}_\mrm{MRC} $    & 81   & 3300          &        $\mathcal{O}(K^2)$ \\ 
\end{tabular}%
}

\label{tables:mj_table}
\end{table}

The vector core consists of 16 single instruction, multiple data lanes and supports different matrix and vector sizes. The processing times for different matrix and vector operations are shown in Table~\ref{tables:mj_table} with $N = 16$ and $K = 16$ or $K = 128$, together with the complexities order of each operation. The inverse of the Gramian is calculated using extended QRD, due to its high performance on this specific vector core\cite{mj}.

The remaining digital logic blocks are implemented on  the AMD ZCU216 FPGA to extract the latency numbers from real hardware. The fronthaul is Ethernet based, and routed directly between panels. The data aggregation is integrated with the Ethernet so that the aggregation can begin immediately when the first data is received. This removes the need for long buffers in relation to the Ethernet. For all hardware units, a clock frequency of $150$MHz is assumed.

\section{Latency Analysis}

\subsection{Latency Model} 

In Fig.~\ref{fig:latency} a break-down of the system latency is displayed. We define the latency as the time starting when the \ac{UE} has new data to transmit and ending when the data for the first subcarrier has been detected by the \ac{BS}, using either \ac{ZF}, \ac{MRC} or \ac{MMSE} algorithms. The latency can be expressed as 
\begin{equation}
\tau=\underbrace{\tau_\mathrm{wait}}_\text{frame}+\underbrace{\tau_\mathrm{LPU}+\tau_\mathrm{fronthaul}+\tau_\mathrm{CPU}}_\text{PHY layer processing}.
\end{equation}
It consists of two main sources, the latency due to the frame structure $\tau_\mathrm{wait}$ and the latency due to the physical layer processing in the \ac{BS}. Based on the architecture in Fig.~\ref{fig:bd}, the \ac{BS} processing latency has three parts $\tau_\mathrm{LPU}$, $\tau_\mathrm{fronthaul}$, and $\tau_\mathrm{CPU}$, which are the latency caused by the \ac{LPU}, fronthaul and \ac{CPU}, respectively. 


The following assumptions are made to simplify latency calculations. Firstly, there are enough hardware resources to match the data throughput of both in-panel processing and inter-panel data transfer. This makes it possible to measure the individual latency of every source in the system without taking the throughput into account (e.g., the extra latency due to buffering can be neglected). Secondly, latency due to the \ac{RF} chain in the \ac{LPU} is known to be very small\cite{rf_latency} and, since it is done in parallel for all antennas, it is ignored. 

The signal propagation, both the one over air between \ac{UE} and \ac{BS}, and the one between panels, scale linearly with distance traversed. Propagation over air takes about $3.3 \mu s$ per km, and, with optical cables used, the total propagation between the panels for a daisy chain topology takes about $5.6 \mu s$ per km of total distance traversed between all panels. As an indoor scenario is considered, the latency caused by these two steps is negligible and not taken into account. Even for an outdoor scenario, the distances need to be in the order of several km for it to have any significant impact. 


When computing $\tau_\mathrm{wait}$, we calculate the longest time from when the \ac{UE} has new data to transmit until there is an available \ac{ULD} slot. This latency depends both on the frame structure and the \ac{OFDM} transmission time. In this analysis, we consider a fixed frame structure as presented in Fig.~\ref{fig:latency}. It is composed of a total of 7 \ac{OFDM} slots, consisting of \ac{UL} and \ac{DL} transmission, as well as a guard time for time-division duplexing switching. The \ac{OFDM} transmission time, including cyclic prefix, is $19 \mu s$, at a subcarrier spacing of 60kHz. The highest $\tau_\mathrm{wait}$ happens when the \ac{UE} gets new data to transmit immediately after the second \ac{ULD} symbol transmission has started, as it then has to wait for six \ac{OFDM} symbols before the next available \ac{ULD} slot.  This makes $\tau_\mathrm{wait} = 133\mu s$. Due to this long wait at the \ac{UE}, any processing necessary at the \ac{UE} side can safely be disregarded, as it will be able to finish within this time frame due to its simplicity. 


For \ac{LPU} the latency, $\tau_\mathrm{LPU}$, is divided into two sources. The latency from the implemented digital front-end and FFT block has a latency of $25\mu s$. It is quite high due to the fact that the full \ac{OFDM} symbol must be received before the computations can be completed. The latency from the local \ac{MIMO} processing is not as straightforward to determine, due to the fact that this latency will depend on both $N$ and $K$. Also, both the processing for the \ac{UL} data symbol and the \ac{UL} pilot need to be taken into account, as the calculations for the \ac{UL} data cannot start until the channel estimate for the relevant subcarrier is available. In \ac{ZF} and \ac{MMSE}, the latency for the local Gramian computation also needs to be taken into account, as it also needs to be transmitted through the fronthaul architecture. The complexity of the relevant calculations are shown in Table~\ref{tables:mj_table}. 

The latency of the Ethernet-based fronthaul, as implemented in the reference architecutre, $\tau_\mathrm{fronthaul}$, grows linearly with the amount of panels according to 

\begin{equation}
    \tau_\mathrm{fronthaul} = 0.87\times (P-1), 
\end{equation}
where $0.87$ is the cost for every interconnection step expressed in $\mu s$. 

For \ac{MRC} $\tau_\mathrm{CPU}$ will be zero, but for \ac{ZF} both the latency of the Gramian inversion and matrix multiplication needs to be taken into account. The complexity of these calculations is also included in Table~\ref{tables:mj_table}. Assuming the central unit knows the \ac{SNR} the only difference between \ac{ZF} and \ac{MMSE} is an addition between the Gramian and a diagonal matrix, which is computed using $K$ additions. As this operation is both of much lower complexity, and consists of additions instead of multiplications, compared to the rest of the central processing, we can ignore the small extra latency this causes and assume that $\tau_\mathrm{CPU}$ for \ac{ZF} and \ac{MMSE} are the same.

\begin{figure}[h]
    \centering 
        \includegraphics[width=1\columnwidth]{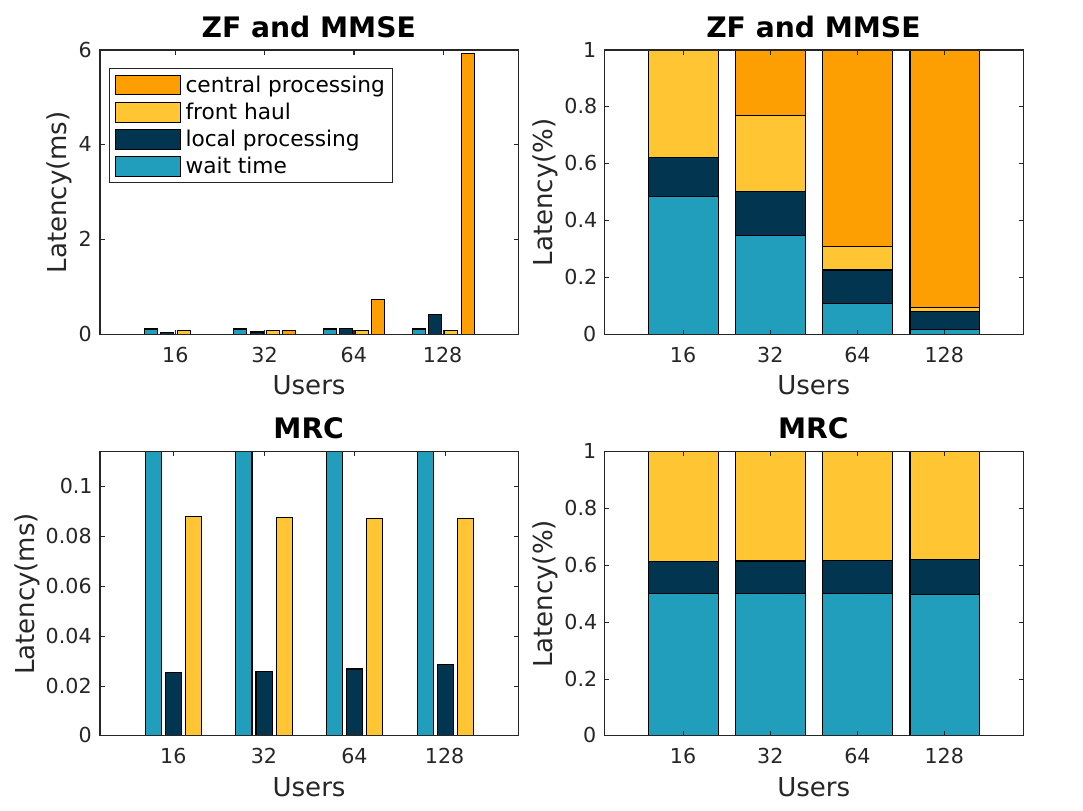}
      \vspace{-2em}
    \caption{The latency, in both absolute and relative values, depending on the amount of \acp{UE}, for a system with 16 antennas per panel and 127 panels.}
 \vspace{-0.2cm}
    \label{fig:barchart}
\end{figure}

\begin{figure*}[h]
     \centering
     \begin{subfigure}
         \centering
         \includegraphics[width=0.9\columnwidth]{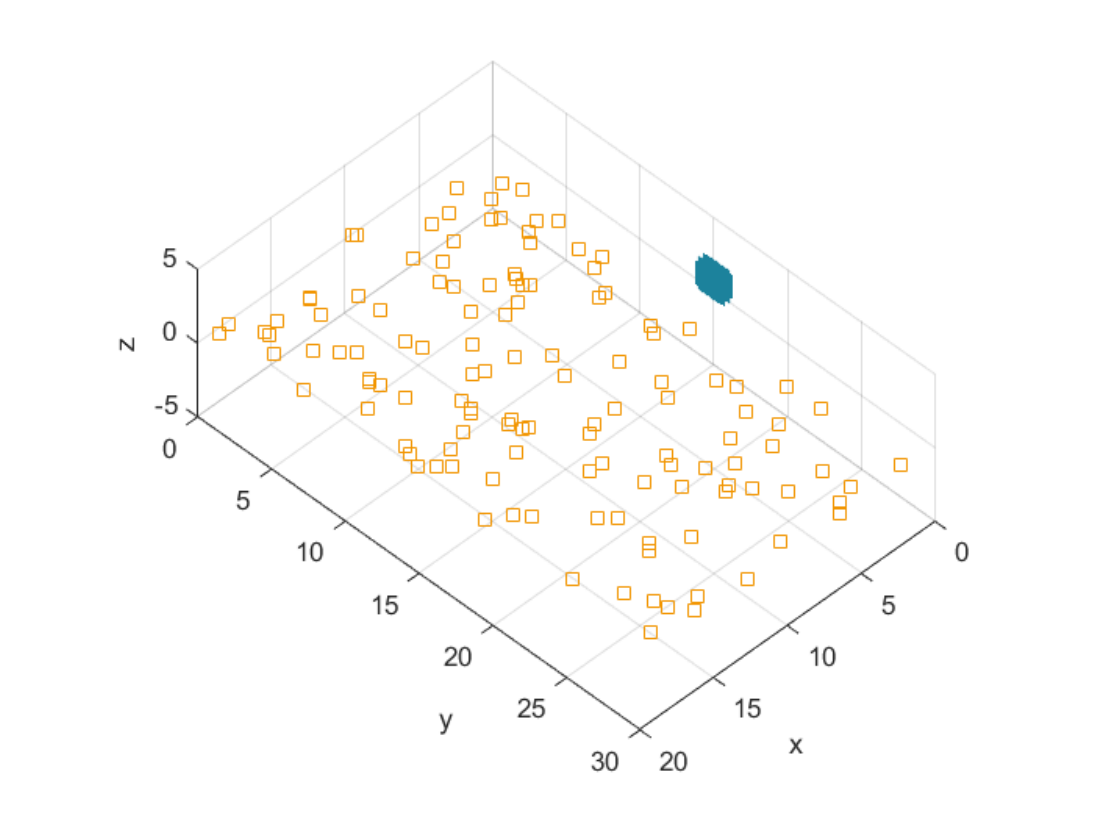}
     \end{subfigure}
     \hfill
     \begin{subfigure}
         \centering
         \includegraphics[width=0.9\columnwidth]{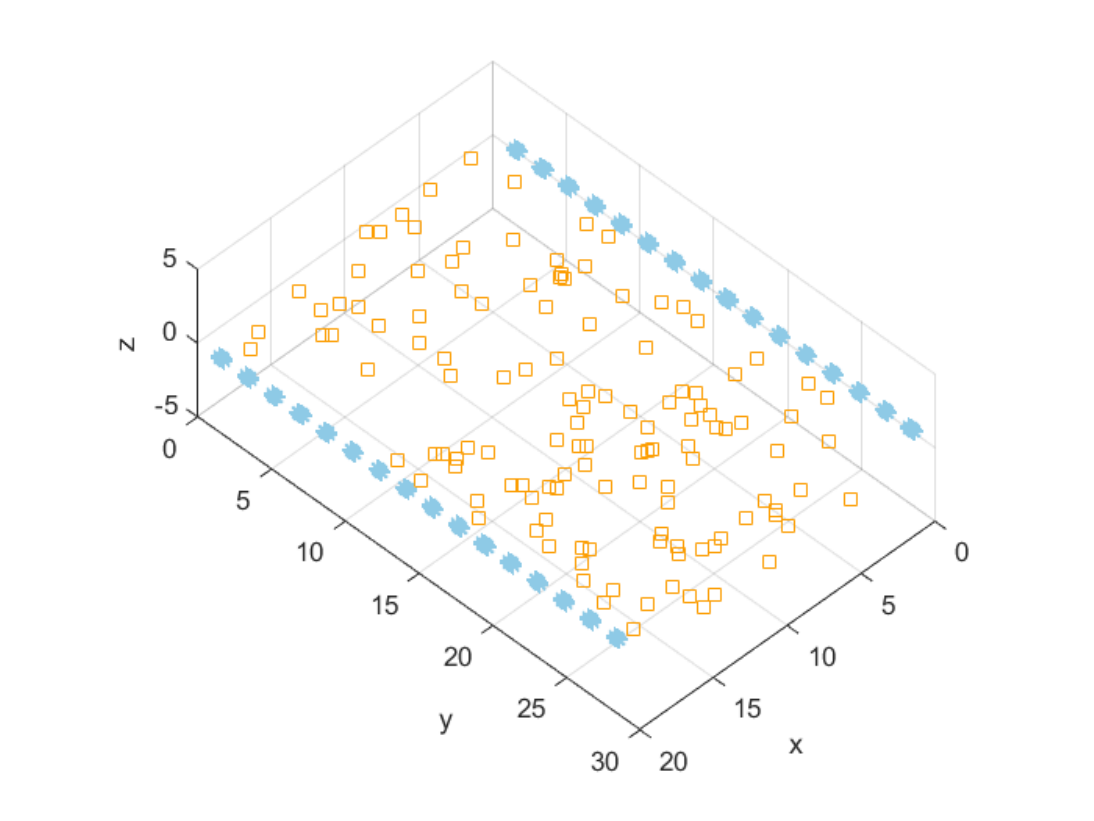}
     \end{subfigure}
     \vspace{-1em}
     \caption{Example scenarios used in the spectral efficiency simulations. $P=1$ with a $32 \times 32$ panel (left) and $P=32$ with $8 \times 4$ panels (right).}
        \label{fig:scen_ex}
        \vspace{-1.5em}
\end{figure*}

\subsection{Latency Results}

In Fig.~\ref{fig:barchart} the latency (absolute and relative) for a system with $N=16$ and $P=128$ is shown for \ac{ZF}, \ac{MMSE} and \ac{MRC}. When the number of \acp{UE} is small, latency distributions are similar, with almost half of the latency due to the frame structure, roughly a third due to the Ethernet-based fronthaul, and the rest caused by the processing. However, when the amount of \acp{UE} grows, two completely different latency ratios appear. For \ac{MRC}, increasing the amount of \acp{UE} does not affect the latency much, as the needed computations only grow linearly with $K$ and are simple to execute. For both \ac{ZF} and \ac{MMSE} though, central processing latency dominates at large numbers of \acp{UE}. While local processing grows quadratically with $K$, the central processing grows cubically with $K$. 


To decrease the latency of a distributed \ac{MIMO} system, using \ac{ZF} or \ac{MMSE}, which is designed to support a large number of of \acp{UE}, further optimizations are clearly needed. A simple solution would be to further optimize the central processing latency, specifically the matrix inversion, though that would likely lead to higher resource and energy costs. Another interesting avenue is the potential of panel segmentation, where only a subset of panels serve certain \acp{UE}. This would lead to smaller matrix sizes and, therefore, simplifying the processing, but it would, in turn, most likely cause new problems.

\section{Latency Versus Spectral Efficiency}

One of the main performance metrics in any wireless communication scenario is the data rate supported by the system. For the considered setting, we are specially interested in the data rates experienced by the different \acp{UE} in the \ac{UL}, how these interplay with the arrangement of the LIS into panels, and the resulting physical layer latency. To this end, we consider an industrial indoor scenario, where the channel matrix
\begin{equation}
\boldsymbol{H}=\sqrt{\frac{K_\mathrm{F}}{1+K_\mathrm{F}}}\boldsymbol{H}_\mathrm{LoS}+\frac{1}{\sqrt{1+K_\mathrm{F}}}\boldsymbol{H}_\mathrm{R},
\end{equation}
is assumed to follow a Rician fading model \cite{paulraj}, with $K_\mathrm{F}$ being the Rician factor, $\boldsymbol{H}_\mathrm{LoS}$ the \ac{LoS} channel between the \acp{UE} and the entire \ac{LIS}, and $\boldsymbol{H}_\mathrm{R}$ an IID Rayleigh fading model where average power for each entry equals the power of the respective \ac{LoS} contribution in $\boldsymbol{H}_\mathrm{LoS}$. The parameters of the scenario are summarized in Table~\ref{tables:sim_scen}. Note that, since we consider \ac{LIS} scenarios with a large number of antenna elements deployed throughout compact indoor regions, we should take into account near-field effects when computing the \ac{LoS} contributions \cite{near-field}. Specifically, we consider that only the perpendicular incident power is absorbed by each antenna element, and we construct the array steering vectors using the spherical propagation model. Moreover, we normalize the channel power such that the SNR value from Table~\ref{tables:sim_scen} corresponds to the average receive SNR considering the whole LIS, where the average is taken over the whole set of simulated scenarios being compared.

\begin{table}[h]
\caption{Scenario parameters.}
\centering
\resizebox{0.85\columnwidth}{!}{%
\begin{tabular}{@{}lc@{}}
\toprule
\textbf{Parameter} & \textbf{Value} \\ 
\midrule
 Frequency & $3.2$ GHz \\ 
  Antenna spacing &  $\lambda/2$\\
   Rician factor & $5$ dB
\end{tabular}%
\hspace{1em} \hspace{1em}
\begin{tabular}{@{}lc@{}}
\toprule
\textbf{Parameter} & \textbf{Value} \\ 
\midrule
SNR & $10$ dB\\
Room length & $30$ m\\
Room width & $20$ m
\end{tabular}%
}

\label{tables:sim_scen}
\end{table}

\begin{figure}
    \centering 
    \includegraphics[width=0.85\columnwidth]{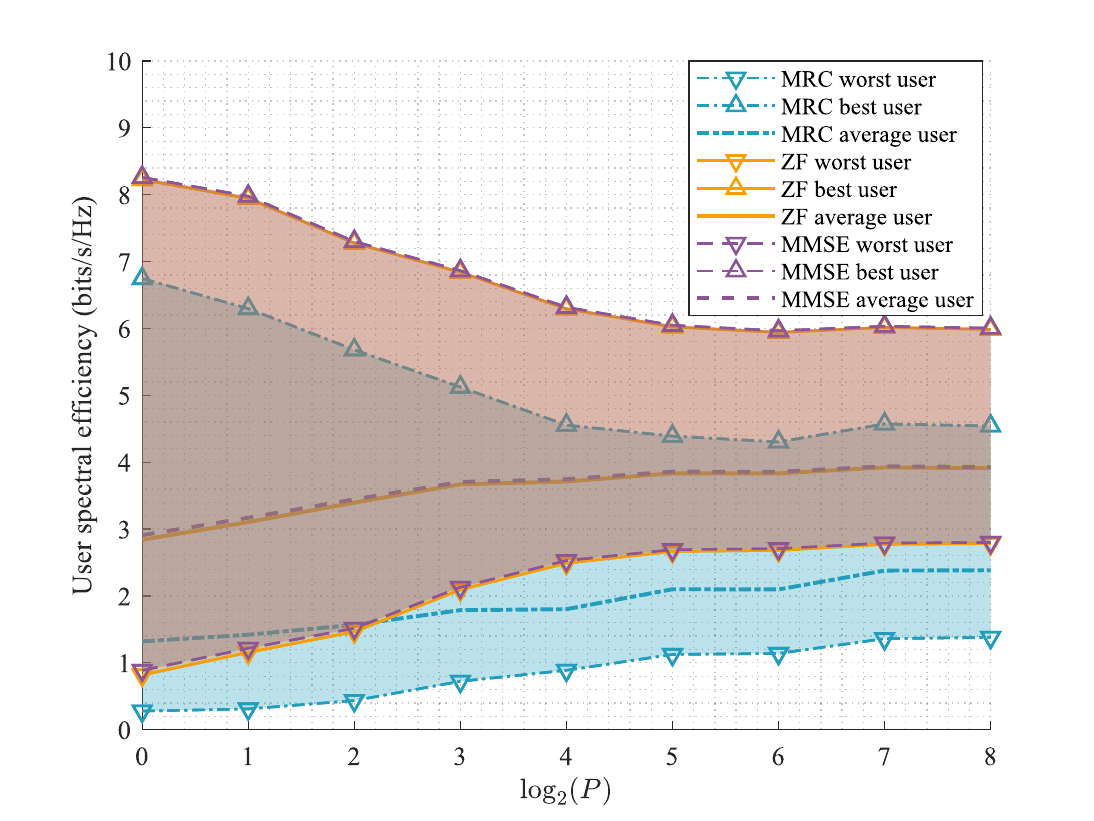}
        \vspace{-0.8em}
    \caption{User spectral efficiencies versus number of panels in logarithmic scale for fixed total number of antennas.}
    \vspace{-1.2em}
    \label{fig:cap_fixedM}
\end{figure}

We begin by analyzing the effect of distributing a fixed number of antennas into a varying number of \ac{LIS}-panels. Specifically, we consider deploying $M=1024$ antennas into equidistant panels along the two walls extending through the length of the room. Note that, in all cases (except when all the antennas are co-located), each panel is facing another panel. The number of panels is incremented from $P=1$ to $P=256$ by powers of 2, while for each $P$ value the respective panels are either square or have twice as many elements along their width than along their height. There are $K=128$ \acp{UE} transmitting equal power, and these are randomly located on a plane at a minimum distance of $10\lambda$ from the walls. The LIS-panels extend from the \acp{UE} plane towards the ceiling of the room. Two example scenarios are shown in Fig.~\ref{fig:scen_ex}. All spectral efficiency results are averaged over $10^2$ realizations of the uniform \ac{UE} placements, each including $10^3$ realizations of the Rayleigh fading channel component. Fig.~\ref{fig:cap_fixedM} plots the spectral efficiencies with respect to the number of panels for the described scenarios. We can see that distributing the antennas throughout the scenario into more smaller \ac{LIS}-panels favors user fairness, reducing potential outages for all processing schemes, while it also allows for a slight increase in average user rate. We believe that this is a strong motivation for favoring more distributed antenna deployments since it increases the overall system reliability. At the same time, as seen in the previous section, in dense user deployments the latency values for the \ac{ZF} and \ac{MMSE} cases are dominated by the central processing required to perform matrix inversion, so the most distributed implementation only has marginal latency increase over having all the antennas co-located in one panel. In the \ac{MRC} case, the fronthaul latency associated to sharing the data sequentially between panels still has noticeable impact, leading to a latency-performance trade-off which should be further analyzed under different settings. Nevertheless, there is a major latency reduction with respect to \ac{ZF} and \ac{MMSE}, at the cost of some degradation in average user spectral efficiency.

\begin{figure}
    \centering 
        \includegraphics[width=0.85\columnwidth]{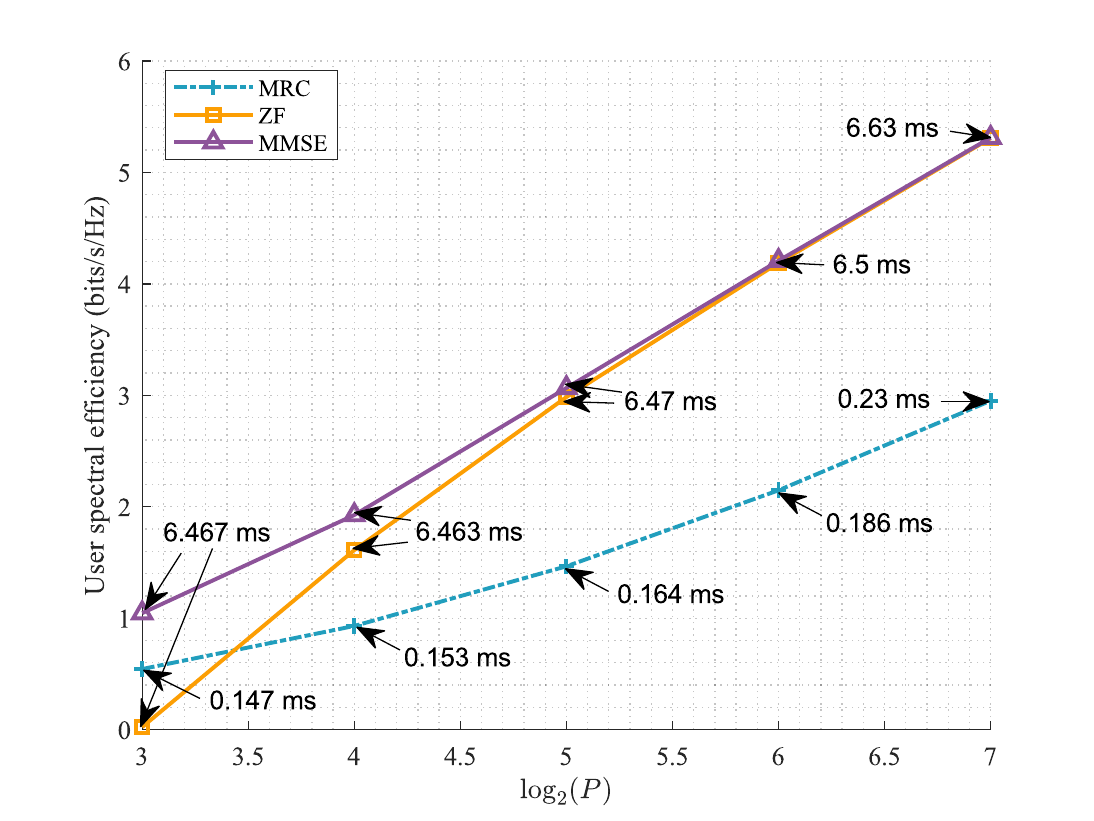}
        \vspace{-0.8em}
    \caption{Average user spectral efficiency versus number of $4\times 4$ panels in logarithmic scale. For each point, the computed latency value is included.}
    \vspace{-1.2em}
    \label{fig:cap_fixedN}
\end{figure}

We now consider the scaling of the spectral efficiencies when deploying varying numbers, $P$, of equally sized square panels with $N=16$ antennas each. Fig.~\ref{fig:cap_fixedN} shows the average user spectral efficiency with respect to the number of panels for the indoor scenario previously described. Aside from the fixed panel size, the same scenario considerations apply as for the results in Fig.~\ref{fig:cap_fixedM}. For large enough $P$, \ac{ZF} and \ac{MMSE} exhibit similar performance, hinting that such scenario is interference limited, while the average user spectral efficiency increases logarithmically with the number of panels, which is justified by the extra array gain associated to having more panels receiving the transmitted power. Fig.~\ref{fig:cap_fixedN} also includes the associated latency values for each panel deployment. Interestingly, the spectral efficiency achieved by \ac{MRC} is comparable to that achieved by \ac{ZF} and \ac{MMSE} with a quarter of the LIS-panels, which still requires a latency close to $30$ times higher. Thus, we conclude that, in the considered scenario, \ac{MRC} is much more effective at exploiting the trade-off between latency and spectral efficiency.

\section{Conclusion}
\label{sec:conclusion}
We have studied the latency in a panel-based LIS indoor scenario with different panel deployments. We presented algorithms for performing decentralized \ac{MRC},  \ac{ZF}  and \ac{MMSE} processing without any loss compared to the centralized counterpart. We also compute latency values based on real implementations for the different schemes, which show that, as the number of users grows, the \ac{ZF} and \ac{MMSE} latency is quickly dominated by the central processing required to invert the channel Gramian. Finally, we compared the spectral efficiency achieved with different deployments, where we observed that having antennas more distributed increases the system reliability, and comes at essentially no extra latency cost when using \ac{ZF} or \ac{MMSE}. However, \ac{MRC} seems to attain a better trade-off between latency and performance.

    \balance
 	\bibliography{references}
    \bibliographystyle{unsrt}
\maketitle

\end{document}